\documentclass[11pt]{article}

\usepackage[preprint]{acl}

\usepackage{times}
\usepackage{latexsym}
\usepackage{subfigure}
\usepackage{multirow}
\usepackage{booktabs}
\usepackage{tabularx}
\usepackage{booktabs}
\usepackage{makecell}

\usepackage{amsmath,amsfonts}
\usepackage{algorithmic}
\usepackage{algorithm}
\usepackage{array}
\usepackage{textcomp}
\usepackage{stfloats}
\usepackage{url}
\usepackage{verbatim}
\usepackage{graphicx}
\usepackage{amsmath}
\usepackage{orcidlink}
\usepackage{tabularx}
\usepackage{marvosym}
\usepackage{subfigure}
\usepackage{multirow}
\usepackage{booktabs}
\usepackage{color}
\usepackage{marvosym}
\usepackage{authblk}
\usepackage[switch]{lineno}

\usepackage[T1]{fontenc}

\usepackage[utf8]{inputenc}

\usepackage{microtype}

\usepackage{inconsolata}

\usepackage{graphicx}

\usepackage{amsmath}
\title{RumorSphere: A Framework for Million-scale Agent-based Dynamic Simulation of Rumor Propagation}

\author[a,c,d]{Yijun Liu}
\author[b,*]{Wu Liu}
\author[a,c,d]{Xiaoyan Gu}
\author[b]{Hantao Yao}
\author[a]{Weiping Wang}
\author[e]{Jiebo Luo}
\author[b]{Yongdong Zhang}

\affil[a]{Institute of Information Engineering, Chinese Academy of Sciences}
\affil[b]{School of Information Science and Technology,
 University of Science and Technology of China}
\affil[c]{School of Cyber Security, University of Chinese Academy of Sciences}
\affil[d]{State Key Laboratory of Cyberspace Security Defense}
\affil[e]{Department of Computer Science, University of Rochester}

\begin{document}
\setlength{\abovedisplayskip}{5pt}
\setlength{\belowdisplayskip}{5pt}
\setlength{\textfloatsep}{18pt}

\maketitle
\begin{abstract}
Rumor propagation modeling is critical for understanding and mitigating misinformation.
Existing approaches combining rule-based regular agents with LLM-driven core agents provide a promising paradigm for large-scale rumor simulation.
However, overlooking the dynamic nature of core agents and the importance of network topology on rumor spread significantly undermines the simulation performance.
To address these issues, we present RumorSphere\footnote{The dataset and source code will be made publicly available after the paper is accepted.}, a dynamic and hierarchical resonance framework for effective rumor simulation at the million-agent scale.
Considering the dynamic role of core agents in rumor evolution, we propose a multi-agent dynamic interaction strategy based on the information cocoon theory, which adaptively identifies and activates critical core agents at conflict boundaries using LLMs, effectively supporting simulations with millions of agents.
In addition, we design a hierarchical resonance network that integrates opinion leaders and localized community structures, enabling more realistic modeling of explosive rumor spread in real-world scenarios.
Experiments on real-world datasets show that RumorSphere outperforms state-of-the-art methods, reducing simulation bias by an average of 26.5\%.
\end{abstract}

\section{Introduction}

With the proliferation of social media, the rapid spread of rumors has emerged as a major global concern \cite{gawronski2024signal,he2024adaptive,xiao2024cross}. 
While extensive research has focused on rumor detection and identification 
\cite{DBLP:conf/acl/TianHHHLQM0L25,DBLP:conf/acl/ZhangLTWW25,DBLP:conf/www/TaoW0W024}, 
modeling their propagation dynamics is equally imperative, as the essence of rumors lies in how they spread and form collective social patterns.
However, rumor propagation is inherently a large-scale and emergent social phenomenon \cite{vosoughi2018spread}, making the accurate modeling of its intricate, time-evolving dynamics a challenging task.

\begin{figure}[!t]
  \centering
  \includegraphics[width=\linewidth]{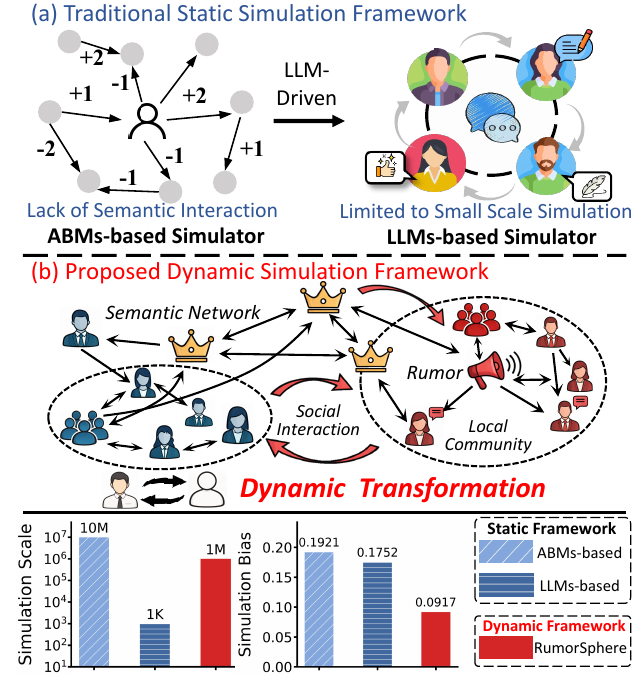}
    \vspace{-5mm}
  \caption{(a) Traditional models are confined by fixed rules and static agent architecture, limiting the effective scalability of the simulation. (b) The proposed RumorSphere adaptively partitions ABM and LLM agents by \textbf{dynamic} interaction strategy for million-scale rumor propagation simulation, significantly reducing the simulation bias.}
  \label{fig:demo}
  \vspace{-5mm}
\end{figure}

Previous research primarily relied on agent-based models (ABMs) to simulate rumor propagation \citep{DBLP:journals/tnsm/XiaoYSL20,DBLP:journals/isci/ZhuYGZ21,DBLP:conf/csse/ZhaoZM20}.
As shown in Fig.~\ref{fig:demo}(a) (left), they are mathematical models that define static agent interaction rules to generate collective social patterns. 
However, such fixed-rule abstractions struggle to capture the bursty, nonlinear, and highly emergent rumor dynamics \cite{gawronski2024signal,vosoughi2018spread}.
In contrast, large language models (LLMs) have demonstrated remarkable capability to mimic human behaviors, with growing research using LLMs to create AI agents for social network simulations \citep{liu2024tiny,DBLP:conf/ijcai/LiuCZG0024,DBLP:conf/naacl/ChuangGHSHYSHR24}, as shown in Fig.~\ref{fig:demo}(a) (right).
Considering that rumor propagation in the real world involves millions of users, many approaches have been proposed to enhance simulation efficiency. 
For instance, OASIS \cite{yang2024oasis} utilizes locally deployed LLMs for parallel inference to scale simulations, but it still requires substantial computational resources. 
Alternatively, some works \citep{DBLP:conf/acl/MouWH24,DBLP:journals/corr/abs-2502-08691} propose to activate only the most influential agents (opinion leaders) as LLMs-driven core agents, with the remainder modeled as efficient ABMs agents. These approaches offer a promising solution for effective large-scale rumor propagation simulation.

However, rumor propagation is a complex emergent phenomenon in which the \textbf{key spreaders driving rumors are dynamically evolving with the environment.}
Due to the transient nature of public opinion, existing approaches that employ a static strategy to activate core agents may cause public opinion to converge with opinion leaders, limiting the effective scalability of the simulation.
Furthermore, these methods often neglect the fact that \textbf{tightly-knit local communities are a key factor in the rapid spread and amplification of rumors}, failing to reflect the explosive and wide-reaching propagation characteristic of real-world rumors.

To handle these challenges, this paper introduces \textbf{RumorSphere}: a novel dynamic and hierarchical resonance framework for rumor propagation simulation at the million-agent scale.
The social media users are simulated as core and regular agents for rumor propagation simulation. 
Considering the dynamic role of core agents in rumor evolution, we propose a Dynamic Interaction Strategy (DIS) based on the information cocoon theory. DIS leverages a real-time information confusion index to adaptively switch agent roles and determine agents' interaction patterns dynamically. 
Agents at ``conflict boundaries'' that bridge opposing communities, exposed to clashing viewpoints and lacking local consensus, are switched to LLM-driven core agents with advanced reasoning capabilities. 
In contrast, massive regular agents, trapped in ``information cocoons'', are modeled using the proposed Confusion-Adaptive Herding (CAH) model. 
This strategy can effectively model large-scale agent behaviors, supporting simulations with millions of agents.
Furthermore, to more realistically model the explosive spread of rumors on large-scale social networks, we introduce a Hierarchical Resonance Network (HRN) based on semantic preferential and triangular connection mechanisms. 
Unlike mechanical topology generators, HRN integrates structural hierarchy with cognitive resonance, modeling how rumors rapidly traverse through influential pathways while being reinforced within tight-knit local communities.
The semantic preferential connection fosters opinion leaders, while the triangular connection facilitates the formation of local communities, thus enhancing the clustering coefficient and reducing the node distance. 
This structure enables faster rumor propagation in the social network, which more closely mirrors real-world rumor dynamics.

Extensive experiments on real-world datasets demonstrate that RumorSphere achieves strong alignment with actual rumor dynamics, significantly outperforming state-of-the-art baselines with an average 26.5\% reduction in simulation bias. Furthermore, it enables effective counterfactual analysis, empirically revealing the importance of cognitive immunization in mitigating rumors.

In summary, the main contributions of this work are as follows:


\begin{itemize}
\item We propose RumorSphere, a pioneering million-scale dynamic rumor propagation simulator based on LLMs. By integrating a dynamic and hierarchical resonance framework with opinion leaders and local community structures, rumors can spread more closely mirror real-world dynamics.
\item We propose a multi-agent dynamic interaction strategy, which can adaptively distinguish core and normal agents at conflict boundaries, effectively supporting simulations with millions of agents.
\item We design the hierarchical resonance network, where the central opinion leaders and localized community structures facilitate more realistic simulations of explosive rumor spread.
\end{itemize}

\begin{figure*}[!t]
  \centering
  \includegraphics[width=\linewidth]{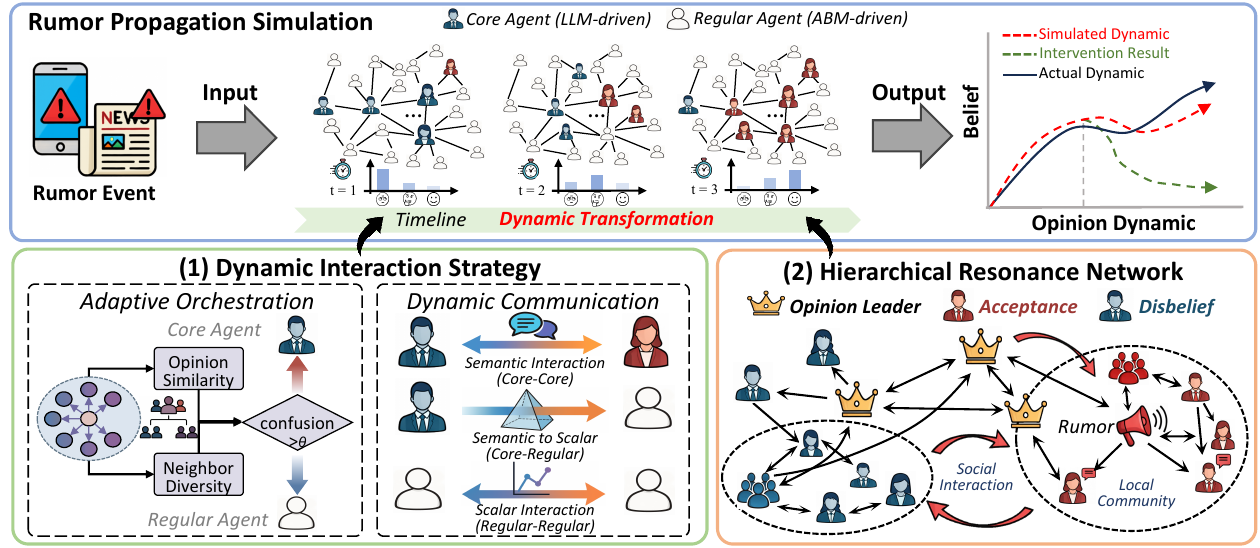}
  \caption{The overview of the RumorSphere framework. The social media users are simulated as core and regular agents for rumor propagation simulation. In the simulation process: (1) Dynamic Interaction Strategy adaptively identifies core agents at conflict boundaries and determines agents' interaction patterns dynamically for driving rumor propagation; and (2) Hierarchical Resonance Network facilitates realistic explosive propagation by integrating opinion leaders with tightly-knit local community structures.}
  \label{fig:overview}
\end{figure*}


\section{Related Work}
\subsection{Rumor Propagation Modeling}
Modeling rumor propagation is crucial for understanding misinformation dynamics and developing countermeasures such as early warning systems \citep{DBLP:conf/ACISicis/HanG18} and information blocking \citep{he2024adaptive}. 
Existing approaches can be broadly categorized into epidemic models, point-process-based models, and agent-based models (ABMs). 
Epidemic models \citep{DBLP:conf/kdd/JinDSCR13, DBLP:journals/amc/HuangCM21, DBLP:journals/tweb/RaponiKOP22}, inspired by infectious disease spread, treat rumor diffusion as a contagion process with individuals transitioning between states based on infection rates (e.g., Susceptible-Infected-Recovered (SIR) model \citep{zhu2017rumor}). Point-process models \citep{DBLP:conf/acl/LukasikCB15, DBLP:conf/dsaa/NieZLDB20, DBLP:conf/cocoa/LiJGY23} capture the temporal randomness of propagation events, modeling the timing and intensity of rumor cascades through random interactions. 
ABMs \citep{DBLP:journals/tnsm/XiaoYSL20, DBLP:journals/tcss/MouXZLX23, DBLP:journals/isci/ZhuYGZ21} employ mathematical formalisms to simulate micro-level information exchange in rumor propagation. 
However, these methods often simplify opinion dynamics into numerical scalars, overlooking the semantic and cognitive complexities essential for accurate rumor modeling. This motivates the use of LLMs to better capture content-driven rumor dynamics.

\subsection{LLM-agent-based Social Network Simulation}
With the rapid advancement of LLMs, LLM-based multi-agent systems have gained significant attention \citep{qian2023communicative, zhou2023sotopia, DBLP:conf/acl/QianDLLXWC0CCL024}. 
The integration of LLM-driven agents into social simulations \citep{DBLP:conf/acl/MouWH24, DBLP:conf/naacl/ChuangGHSHYSHR24, liu2024tiny} represents an emerging research field with promising results. 
For example, \citep{tornberg2023simulating} used generative agents to explore social media's impact on polarization, while \citep{DBLP:conf/uist/ParkOCMLB23} demonstrated that LLM-driven simulacra can replicate human-like social interactions. 
In misinformation studies, \citep{liu2024tiny} utilized LLMs to model the lifecycle of fake news, and \citep{DBLP:conf/ijcai/LiuCZG0024} examined rumor propagation dynamics in small communities to explore suppression strategies. 
Some approaches \citep{DBLP:conf/acl/MouWH24, yang2024oasis, DBLP:journals/corr/abs-2502-08691} scale simulations to 100,000 agents by activating a fixed subset of core agents, demonstrating the feasibility of LLM-driven simulation at scale.
However, these static strategies fail to capture the evolving nature and the bursty patterns of rumor propagation.
In this work, we dynamically orchestrate agents and construct the HRN to effectively model million-scale dynamic rumor propagation.




\section{Methodology}

\subsection{Simulation Framework}
As shown in Fig.~\ref{fig:overview}, RumorSphere operates as a social network sandbox, modeling users as agents to simulate emergent rumor dynamics. 
The simulation workflow is orchestrated by the Dynamic Interaction Strategy (DIS). 
Specifically, at each round shown in Fig.~\ref{fig:overview} (1), the DIS quantifies a real-time ``information confusion'' index for each agent to dynamically orchestrate agent behaviors: pivotal nodes are designated as core agents driven by LLM, while the majority operate as regular agents governed by the efficient Confusion-Adaptive Herding (CAH) model. 
As shown in Fig.~\ref{fig:overview} (2), these heterogeneous agent interactions occur within an Hierarchical Resonance Network (HRN) topology, which configures the underlying connectivity to support large-scale, explosive rumor propagation. 
Finally, the continuous interactions among these heterogeneous agents generate the evolving rumor dynamics over time.

\subsection{Core Agent}
\label{sec:core}
To model pivotal users who drive the narrative evolution of rumors, we design core agents powered by LLMs. Unlike regular agents, these agents utilize specialized persona, memory, and action modules to cognitively process rumor information and execute complex propagation behaviors.

\subsubsection{Persona Module}
The basic persona encompasses demographic attributes, including name, gender, occupation, interests, and personality traits \cite{DBLP:conf/hicss/BrunkerWMM20}. 
Crucially, these attributes determine the agent's baseline susceptibility and inherent stance toward specific rumor events. 
To ensure the fidelity of the simulated population, we initialize these attributes using a hybrid approach that synergizes sampling from real-world datasets with stochastic assignment. Detailed initialization processes are provided in the Appendix.

\subsubsection{Memory Module}
An individual's belief evolution is shaped by inherent traits and cumulative environmental exposure \cite{DBLP:journals/advcs/DeffuantNAW00}. To model this, we use a dual-memory architecture: \textbf{Personal experience} records the agent's historical stance and behaviors, while \textbf{environmental memory} tracks peer interactions and trending rumors, simulating cognitive effects of the social context. 
Before interaction, agents retrieve relevant memories to guide decision-making. Afterward, new observations are stored in both natural language and vector embeddings for future use. Details of the memory mechanism are provided in the Appendix.

\subsubsection{Action Module}
Inspired by the Twitter environment and \citep{DBLP:conf/acl/MouWH24}, we design an action module where discrete actions serve as the fundamental mechanism for rumor diffusion. The actions include: (1) \textbf{Post}: Initiating original content to spread or debunk rumors; (2) \textbf{Retweet}: Amplifying existing rumor cascades; (3) \textbf{Reply}: Debating or reinforcing viewpoints; (4) \textbf{Like}: Approving tweets; (5) \textbf{Do Nothing}: Remaining silent. Optional actions are presented via prompts, and the agent's responses are parsed to reflect their specific impact on the environment. Examples of agents' decision-making are presented in the Appendix.

\subsection{Regular Agent}
\label{sec:ordinary}
To simulate the massive population of regular agents, we model them using efficient ABMs.
However, traditional ABMs typically rely on static interaction parameters, failing to capture the dynamic psychology of uncertainty-driven herding during rumor outbreaks. 
To address this, we propose the Confusion-Adaptive Herding (CAH) model, which integrates cognitive states with social influence, reflecting the sociological insight that users tend to seek consensus amidst confusion and demonstrate obedience to authority \cite{sunstein2014rumors}. 
In this model, each agent maintains a continuous opinion score $o_{i,t} \in [-1, 1]$, ranging from total disbelief ($-1$) to full acceptance ($1$). 
The interaction dynamics are governed by three functions: $f_{select}$ filters neighbors to identify eligible influencers based on bounded confidence; $f_{update}$ updates opinion scores based on the influencers and the agent's confusion level regarding the rumor; and $f_{message}$ determines the specific rumor signal transmitted to neighbors. 
Details are provided in the Appendix.

\subsection{Dynamic Interaction Strategy}
\label{sec:dis}
To dynamically identify the core agents during rumor propagation and orchestrate heterogeneous agent behaviors, we propose a multi-agent DIS. 
Unlike prior works relying on the static assignment of core agents, DIS dynamically empowers agents situated at evolving conflict boundaries—critical junctions where opposing rumor narratives collide.
To achieve this, the DIS includes two synergistic components: Adaptive Orchestration (AO) for dynamic role assignment and Dynamic Communication (DC) for managing heterogeneous interactions.

\subsubsection{Adaptive Orchestration}
To quantify the intensity of cognitive conflict experienced by an agent, we introduce the information confusion index ($\tau_{i,t}$). 
The rationale lies in the structural distinction of social roles: agents within echo chambers experience high cognitive resonance (homogeneity), whereas agents at conflict boundaries bridge opposing communities, facing clashing viewpoints (high diversity) while lacking local consensus (low similarity).
We capture this by defining opinion similarity $s_{i,t}$ and neighbor diversity $d_{i,t}$ metrics.

$s_{i,t}$ measures the conformity between an agent's opinion $o_{i,t}$ and its local neighborhood $\mathcal{N}_i$:
\begin{equation}
    s_{i,t} = \frac{1}{\left|\mathcal{N}_i\right|} \sum_{j \in \mathcal{N}_i}\left(1-\frac{\left|o_{j,t}-o_{i,t}\right|}{2}\right).
\end{equation}

$d_{i,t}$ quantifies the variance or heterogeneity of opinions within the neighborhood:
\begin{equation}
    d_{i,t} = \sqrt{\frac{1}{\left|\mathcal{N}_i\right|} \sum_{j \in \mathcal{N}_i}\left(o_{j,t}-\mu_{i,t}\right)^2},
\end{equation}
where $\mu_{i,t} = \frac{1}{\left|\mathcal{N}_i\right|} \sum_{j \in \mathcal{N}_i}o_{j,t}$ represents the local consensus.

Accordingly, we formulate the information confusion index $\tau_{i,t}$ as:

\begin{equation}
\label{eq:tau}
    \tau_{i,t} =  2(1 - s_{i,t}) \cdot d_{i,t},
\end{equation}
The factor 2 ensures normalization to $[0, 1]$, as maximum neighborhood diversity ($d_{i,t} \to 1$) inherently limits opinion similarity to 0.5.

Finally, agents exceeding a confusion threshold ($\tau_{i,t} > \theta$) and minimum topological influence (neighbor count) are identified as core agents. This strategy effectively concentrates high-cost reasoning resources on pivotal nodes that determine the trajectory of rumor evolution.

\subsubsection{Dynamic Communication}
The DC module modulates information flow across the heterogeneous population by adapting communication protocols: 
\textbf{Semantic Interaction (Core-Core):} Core agents engage in high-fidelity dialogue using natural language generated by LLMs, allowing for complex argumentation and persuasion. 
\textbf{Semantics-to-Scalar (Core-Regular):} To bridge the modality gap, the LLM-generated text is quantified into an opinion score via an LLM (example and accuracy validation in the Appendix), translating semantic influence into a format compatible with the CAH model.
\textbf{Scalar Interaction (Regular-Regular):} Regular agents exchange continuous opinion scalars via the CAH mechanism, simulating the subconscious assimilation typical of simple contagion processes.

\subsection{Hierarchical Resonance Network}
\label{sec:network}
To more realistically model the explosive spread of rumors, we introduce the Hierarchical Resonance Network (HRN).
Unlike mechanical topology generators, HRN integrates structural hierarchy with cognitive resonance, modeling how rumors rapidly traverse through influential pathways while being reinforced within tight-knit local communities.

\textit{Network Initialization}:
The construction initiates with a fully-connected seed graph $G_0 = (V_0, E_0)$, where $V_0=\{v_1,...,v_n \mid n \ll N\}$ and the edge set satisfies:
\begin{equation}
    \forall v_i, v_j \in V_0, \quad e_{i j} \in E_0 \Longleftrightarrow e_{j i} \in E_0.
\end{equation}

\textit{Iterative Evolution}:
At each step, a new agent node $v_t$ with a specific persona embedding $\mathbf{h}_t$ joins the network. It establishes $m$ connections based on a probabilistic mixture of semantic preferential and triangular  connections:
\begin{equation}
P_{\text{attach}} = 
\begin{cases}
P_{\text{pref}}, & \text{with probability } p \\ 
P_{\text{tri}}, & \text{with probability } (1-p)
\end{cases}.
\end{equation}

\textit{Semantic Preferential Connection ($P_{\text{pref}}$)}:
This mechanism models the ``rich-get-richer'' phenomenon modulated by cognitive homophily. The probability that $v_t$ connects to an existing node $v_i$ is determined by:
\begin{equation}
    P_{\text{pref}}\left(v_t \rightarrow v_i\right) = \frac{k_i \cdot \text{Sim}(\mathbf{h}_t, \mathbf{h}_i)}{\sum_{j \in V_{t-1}} k_j \cdot \text{Sim}(\mathbf{h}_t, \mathbf{h}_j)},
\end{equation}
where $k_i$ (degree) represents structural authority to promote opinion leader emergence, and $\text{Sim}(\cdot)$ denotes the cosine similarity of persona embeddings. The term $\text{Sim}(\cdot)$ ensures cognitive resonance between leaders and followers.

\begin{table*}[ht]
\resizebox{\textwidth}{!}{%
\begin{tabular}{l|llll|llll|llll}
\hline
\multicolumn{1}{c|}{\multirow{2}{*}{\textbf{Method}}} & \multicolumn{4}{c|}{\textbf{Moon Landing Conspiracy}}                                                                        & \multicolumn{4}{c|}{\textbf{Xinjiang Cotton}}                                                              & \multicolumn{4}{c}{\textbf{Trump-Russia Connection}}                                                       \\ \cline{2-13} 
\multicolumn{1}{c|}{}                                 & \multicolumn{1}{c}{$\Delta$$Bias$↓} & \multicolumn{1}{c}{$\Delta$$Div$↓} & \multicolumn{1}{c}{$DTW$↓} & \multicolumn{1}{c|}{$Corr$↑} & \multicolumn{1}{c}{$\Delta$$Bias$↓} & \multicolumn{1}{c}{$\Delta$$Div$↓} & \multicolumn{1}{c}{$DTW$↓} & \multicolumn{1}{c|}{$Corr$↑} & \multicolumn{1}{c}{$\Delta$$Bias$↓} & \multicolumn{1}{c}{$\Delta$$Div$↓} & \multicolumn{1}{c}{$DTW$↓} & \multicolumn{1}{c}{$Corr$↑} \\ \toprule
Deffuant \citep{DBLP:journals/advcs/DeffuantNAW00}                                              & \underline{0.1211}                          & 0.1641                         & 0.5803                        & 0.1737                          & 0.2179                          & 0.1643                         & 0.9118                        & 0.7716                          & 0.2366                          & 0.1883                         & 0.6951                        & -0.1721                         \\
BCMM \citep{DBLP:journals/jasss/HegselmannK02}                                          & 0.1357                          & 0.1535                         & \underline{0.3799}                        & -0.2340                          & 0.4724                          & 0.3047                         & 1.8559                        & -0.7409                          & 0.1760                          & 0.1911                         & 0.5661                        & -0.2090                         \\ 
RA \citep{deffuant2002can}                                          & 0.1923                          & 0.1541                        & 0.6138                        & -0.2482                          & 0.2839                          & 0.1570                         & 0.7324                        & 0.7871                          & \underline{0.1001}                          & \textbf{0.0644}                         & 0.3920                        & \underline{0.7139}                         \\ 
SJM \citep{DBLP:journals/cmot/JagerA05}                                          & 0.2927                          & 0.1138                         & 0.9603                        & \textbf{0.5503}                          & 0.1798                          & 0.1767                         & 0.7524                        & 0.6581                          & 0.1973                          & 0.2252                         & 0.5026                        & -0.5097                         \\ 
Lorenz \citep{lorenz2021individual}                                          & 0.2494                          & 0.1736                         & 0.9647                        & -0.5067                          & 0.4726                          & 0.2737                         & 1.8865                        & -0.9407                          & 0.3305                          & 0.1576                         & 1.2135                        & 0.0740                         \\ 
\midrule
FPS \citep{DBLP:conf/ijcai/LiuCZG0024}                                                  & 0.3088                          & \underline{0.1015}                         & 1.0703                        & -0.1417                          & 0.5580                          & 0.2690                         & 2.0019                        & -0.4615                          & 0.4273                          & 0.1584                         & 1.4522                        & -0.3639                         \\
SOD \citep{DBLP:conf/naacl/ChuangGHSHYSHR24}                                                   & 0.2343                          & 0.1471                         & 0.8826                        & -0.4129                          & 0.2730                          & 0.1651                         & 0.9695                        & 0.8919                          & 0.1199                          & 0.0764                         & \underline{0.3839}                        & 0.3700                         \\
HiSim \citep{DBLP:conf/acl/MouWH24}                                                 & 0.1745                     & 0.2183                    & 0.6863                   & 0.2672                    & \underline{0.1532}                          & \underline{0.0958}                         & \underline{0.3912}                        & \underline{0.9349}                          & 0.1979                          & 0.2113                         & 0.8137                        & 0.0306                         \\
\midrule
\textbf{RumorSphere}                                   & \textbf{0.1048}            & \textbf{0.0931}           & \textbf{0.3751}          & \underline{0.5367}                    & \textbf{0.0761}                          & \textbf{0.0852}                         & \textbf{0.2485}                        & \textbf{0.9373}                          & \textbf{0.0942}                          & \underline{0.1048}                         & \textbf{0.2537}                        & \textbf{0.7487}                         \\ \hline
\end{tabular}%
}
\caption{Comparison of simulated rumor dynamics against real-world data. Results are averaged over 10 runs. \textbf{Bold} and \underline{underlined} indicate the best and second-best performance. ``$\downarrow$''/``$\uparrow$'' indicate that lower/higher is better.}
\label{tab:com}
\end{table*}

\begin{figure*}[ht]
  \centering
    \subfigure[Moon Landing Conspiracy Event]{\includegraphics[width=0.295\linewidth]{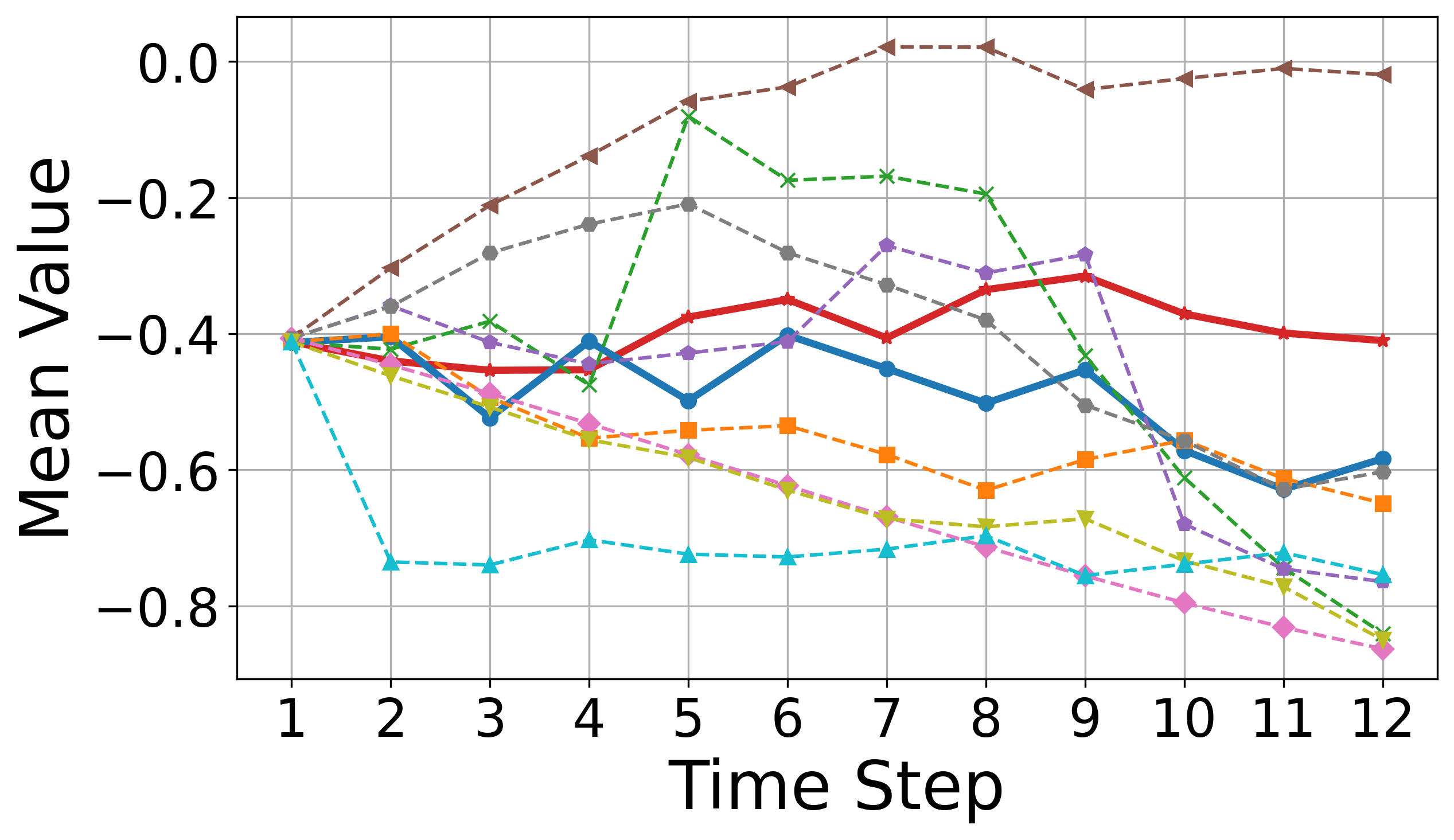}}
    \subfigure[Xinjiang Cotton Event]{\includegraphics[width=0.295\linewidth]{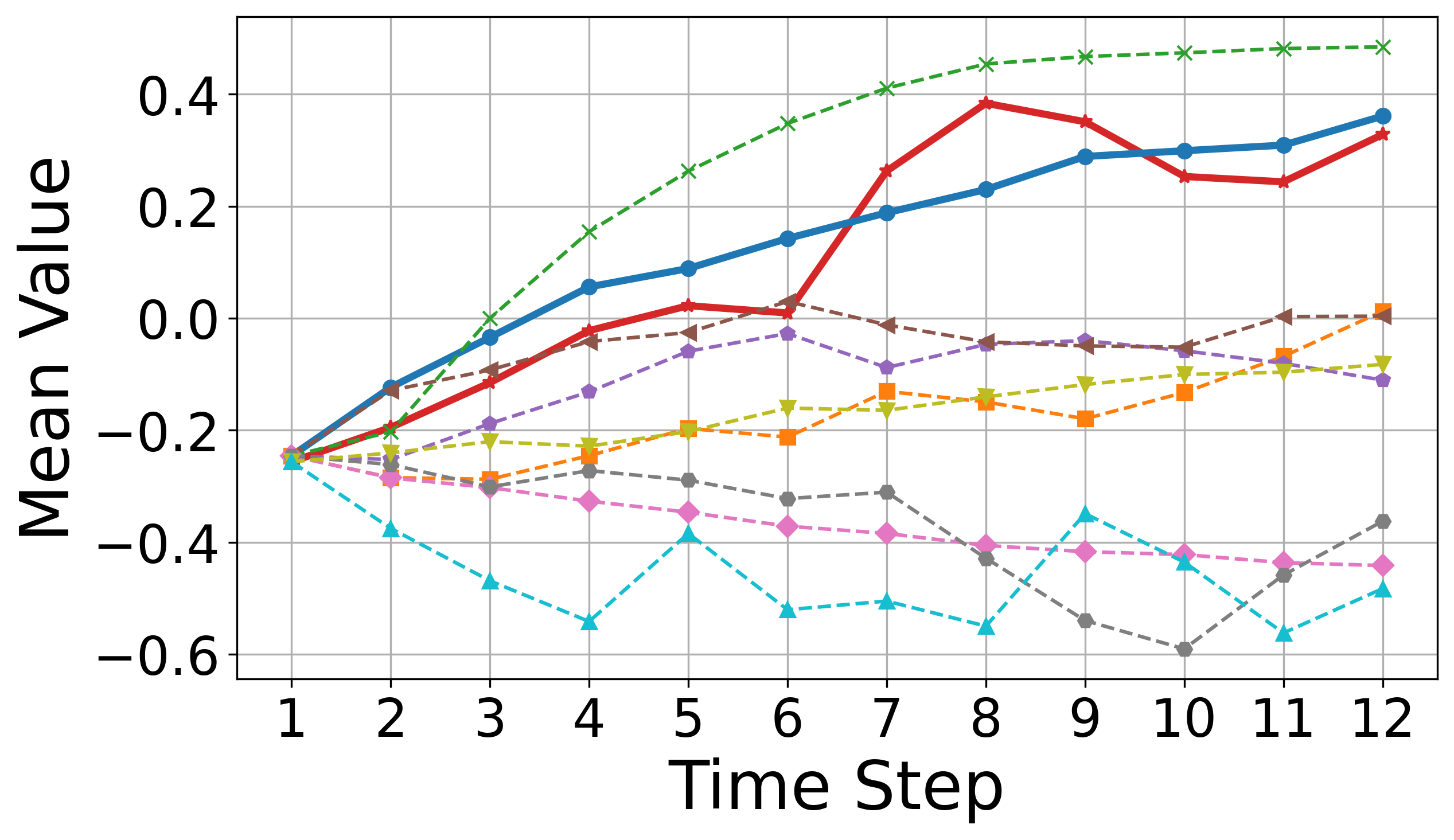}}
    \subfigure[Trump-Russia Connection Event]    
    {\includegraphics[width=0.39\linewidth]{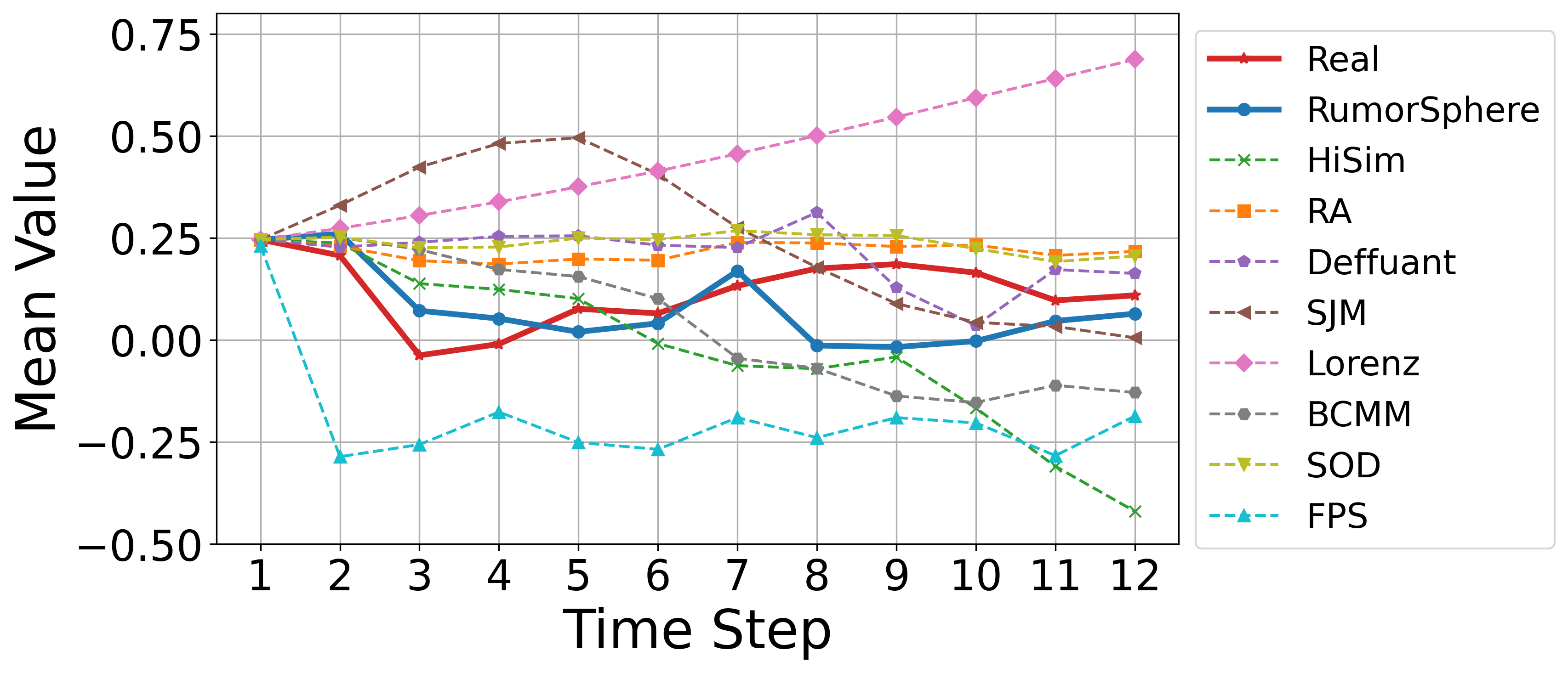}}
  \caption{Visual comparison of temporal opinion evolution across three real-world rumor events. The plots illustrate the trajectories generated by RumorSphere and state-of-the-art baselines against the ground truth.}
  
  \label{fig:trend}
\end{figure*}

\textit{Triangular  Connection ($P_{\text{tri}}$)}:
To simulate the formation of ``local communities,'' we employ a triadic closure process. First, $v_t$ selects a node $v_i$ uniformly at random:
\begin{equation}
    P_{\text{rand}}\left(v_t \rightarrow v_i\right) = \frac{1}{|V_{t-1}|}.
\end{equation}
Subsequently, $v_t$ connects to a neighbor $v_j \in \mathcal{N}(v_i)$ with conditional probability:
\begin{equation}
    P_{\text{tri}}\left(v_t \rightarrow v_j \mid v_t \rightarrow v_i\right) = \frac{1}{|\mathcal{N}(v_i)|}.
\end{equation}
This mechanism significantly boosts the local clustering coefficient, creating dense communities where rumors are repeatedly reinforced among trusted peers.

Through iteration, the HRN evolves into a hierarchical topology characterized by a heavy-tailed degree distribution. This structure supports both the explosive spread of rumors via influential hubs and their persistence within tight communities.

\section{Experiments}
\subsection{Configuration}
We employ ChatGPT (version: gpt-4o-mini) as the backbone for core agents, with a generation temperature of $1.0$ to ensure output diversity. The orchestration threshold is set to $\theta=0.6$, and the minimum topological influence is defined as $\min(0.01N, 1000)$, where $N$ represents the simulation scale. 
More parameter analysis are provided in the Appendix.




\subsection{Datasets}
We validate our model using three high-profile real-world datasets collected from Twitter: the Moon Landing Conspiracy, Xinjiang Cotton, and the Trump-Russia Connection. 
These datasets cover diverse domains and collectively comprise over 30,000 tweets, with temporal spans exceeding six months.
Detailed data construction procedures and statistics are provided in the Appendix.


\subsection{Metrics}
Following previous works \cite{DBLP:conf/acl/MouWH24,DBLP:conf/ijcai/LiuCZG0024}, we evaluate simulation effectiveness by comparing simulated public opinion with real-world data. We use $\Delta Bias$ and $\Delta Div$ for deviation and stability, and apply Dynamic Time Warping ($DTW$) and Pearson correlation ($Corr$) for temporal alignment. We also assess polarization and the echo chambers phenomenon using $P_z$, Global Disagreement ($GD$), and Normalized Clustering Index ($NCI$). Details are in the Appendix.

\subsection{Baselines} 
To comprehensively evaluate the performance of RumorSphere, we benchmark it against two categories of state-of-the-art methods: (1) Traditional ABMs, including Deffuant~\cite{DBLP:journals/advcs/DeffuantNAW00}, BCMM~\cite{DBLP:journals/jasss/HegselmannK02}, RA~\cite{deffuant2002can}, SJM~\cite{DBLP:journals/cmot/JagerA05}, and Lorenz~\cite{lorenz2021individual}, which simulate rumor dynamics through mathematical interaction rules; and (2) LLM-based Multi-Agent Systems, specifically FPS~\cite{DBLP:conf/ijcai/LiuCZG0024}, SOD~\cite{DBLP:conf/naacl/ChuangGHSHYSHR24}, and HiSim~\cite{DBLP:conf/acl/MouWH24}, which leverage LLM to drive complex agent behaviors and simulate rumor dynamics. Detailed model implementations are provided in the Appendix.

\subsection{Main Results}
As shown in Table~\ref{tab:com}, RumorSphere consistently outperforms state-of-the-art baselines, reducing average $\Delta Bias$ and $DTW$ by 26.5\% and 24.1\%. 
RumorSphere also exhibits exceptional adaptability, accurately replicating diverse rumor patterns, from sustained skepticism in Moon Landing to explosive escalation in Xinjiang Cotton, where it reduces simulation bias by over 50.3\% compared to the best baseline.
Notably, RumorSphere achieves an average $Corr$ of 0.7409, which demonstrates its strong temporal synchronization with real-world rumor dynamics.
Furthermore, as visualized in Fig.~\ref{fig:trend}, RumorSphere maintains higher alignment with real-world dynamics than static baseline models, validating its robustness for capturing large-scale, sustained rumor propagation.

\begin{table}[t]
\centering
\resizebox{\columnwidth}{!}{%
\begin{tabular}{l|cccc}
\toprule
\textbf{Model}      & \textbf{$\Delta$Bias$\downarrow$}   
                    & \textbf{$\Delta$Div$\downarrow$}    
                    & \textbf{$DTW$\,$\downarrow$}        
                    & \textbf{$Corr$\,$\uparrow$}   \\
\midrule
\textbf{RumorSphere} & \textbf{0.0761} & \textbf{0.0852} & \textbf{0.2485} & \textbf{0.9373} \\
\midrule
-w/o AO-static       & 0.1361          & 0.1257          & 0.4529          & 0.8964 \\
-w/o AO-random       & 0.2539          & 0.1792          & 0.9916          & 0.8882 \\
\midrule
-w/o HRN-regular     & 0.3440          & 0.2127          & 1.3943          & 0.7616 \\
-w/o HRN-random      & 0.2938          & 0.1909          & 1.1838          & 0.9215 \\
-w/o HRN-BA & 0.1832          & 0.1725          & 0.5633          & 0.8431 \\
-w/o HRN-small-world & 0.2416          & 0.1559          & 0.6342          & 0.8754 \\
\midrule
-w/o CAH-Deffuant    & 0.1421          & 0.0951          & 0.4151          & 0.8994 \\
-w/o CAH-RA          & 0.1632          & 0.1271          & 0.5269          & 0.7315 \\
\midrule
-Smaller scale (100) & 0.2491          & 0.1337          & 0.8145          & 0.9307 \\
\bottomrule
\end{tabular}
}
\caption{Ablation results on the Xinjiang Cotton dataset. ``w/o X-Y'' indicates replacing component X (e.g., AO, HRN, CAH) with baseline Y.}
\label{tab:abl}
\end{table}

\begin{figure}[t]
  \centering
    \subfigure[Token consumption under varying simulation epochs]{\includegraphics[width=0.49\linewidth]{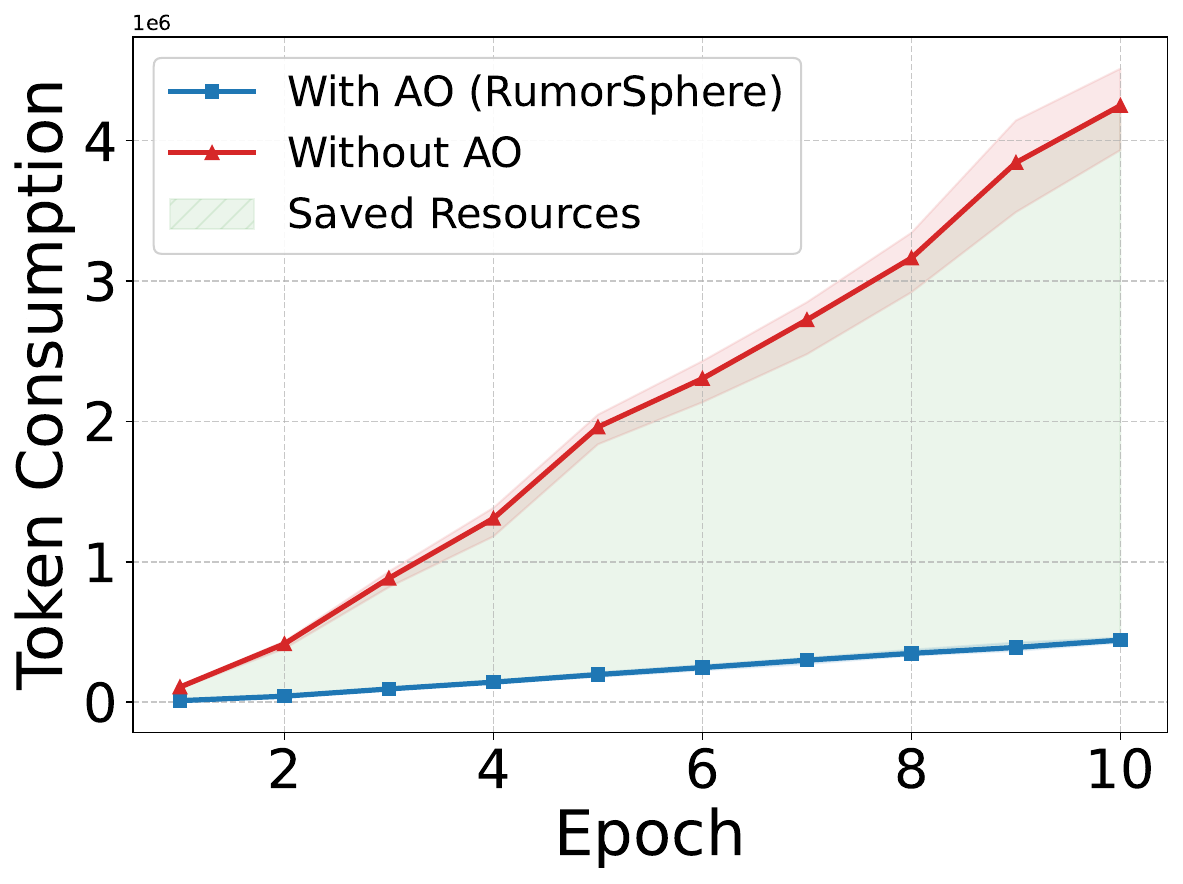}}
    \subfigure[Token consumption under varying agent scales]{\includegraphics[width=0.49\linewidth]{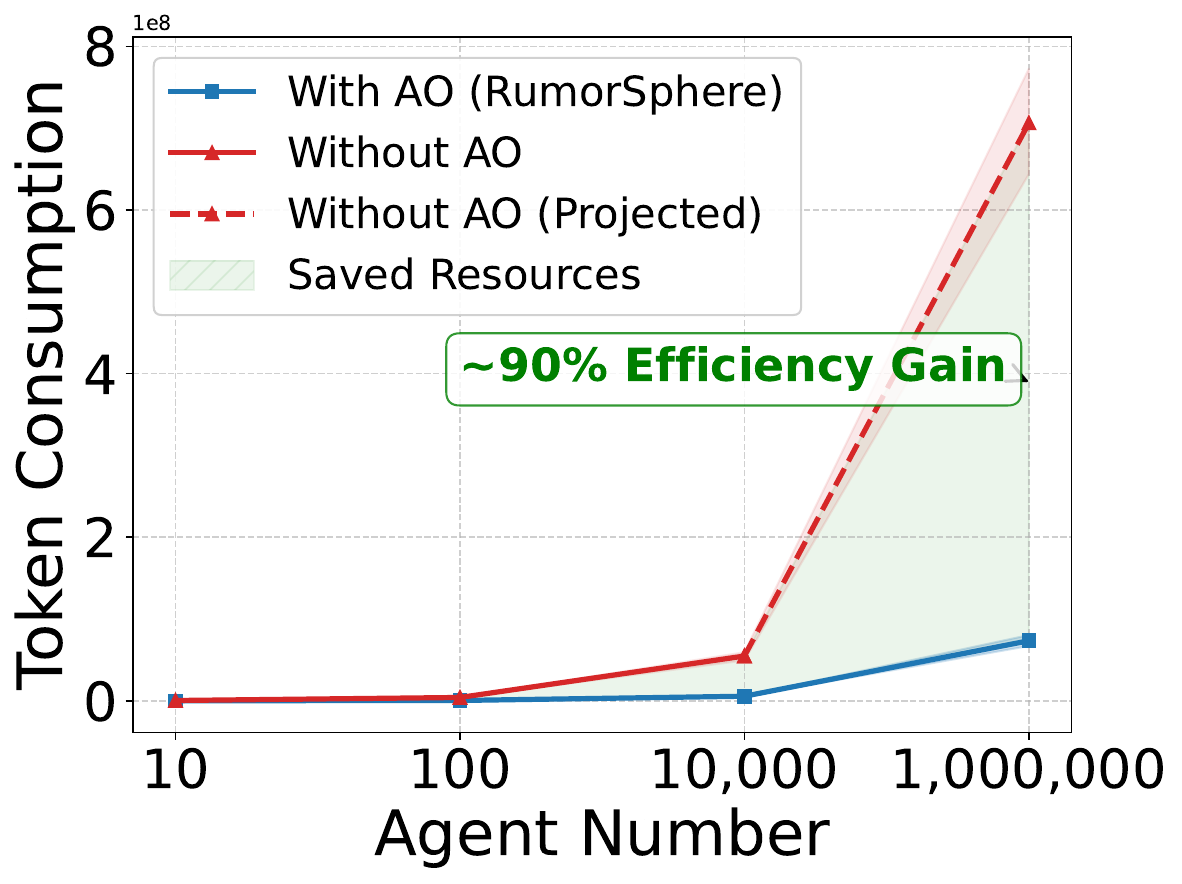}}
  \caption{Token consumption across different scenarios.}
  \label{fig:token}
\end{figure}


\subsection{Ablation Study}
We validate the effectiveness of our core components, AO, HRN, and CAH, through ablation experiments on the Xinjiang Cotton dataset.

\textbf{The Impact of AO:} As shown in Table~\ref{tab:abl}, removing AO leads to a notable decline in simulation fidelity. Specifically, the ``Random'' strategy increases $\Delta Bias$ by 0.181 due to resource wastage on non-critical agents, while the ``Static'' strategy causes unnatural opinion convergence dominated by initial influencers. 
Moreover, Fig.~\ref{fig:token} demonstrates that AO reduces token consumption tenfold compared to full-LLM approaches. This validates that AO effectively balances behavioral realism with computational efficiency, enabling million-scale simulations.
Detailed analysis of the temporal dynamics regarding the number of activated core agents is provided in the Appendix.

\begin{table}[t]
    \centering
    \label{tab:topology}
    \renewcommand{\arraystretch}{1.1}
    \resizebox{\columnwidth}{!}{
        \begin{tabular}{lccc}
            \toprule
            \textbf{Model} & \textbf{Exponent} & \textbf{Clustering} & \textbf{Path} \\
            \midrule
            \textit{Typical Social Nets} & $2.0 \sim 3.0$ & High ($>0.1$) & Low \\
            \midrule
            Regular Lattice & N/A$^{\dagger}$ & 0.667 & 500.45 \\
            Random Graph & N/A$^{\dagger}$ & 0.001 & 4.25 \\
            BA Model & 2.97 & 0.007 & 3.67 \\
            Small-world Model & N/A$^{\dagger}$ & 0.489 & 4.15 \\
            \textbf{HRN (Ours)} & 2.38 & 0.163 & 3.72 \\
            \bottomrule
            \multicolumn{4}{l}{\footnotesize $^{\dagger}$Distribution does not follow a power-law.}
        \end{tabular}
    }
\caption{Topological comparison of generated networks.}
\end{table}
\begin{figure}[t]
  \centering
  \includegraphics[width=\linewidth]{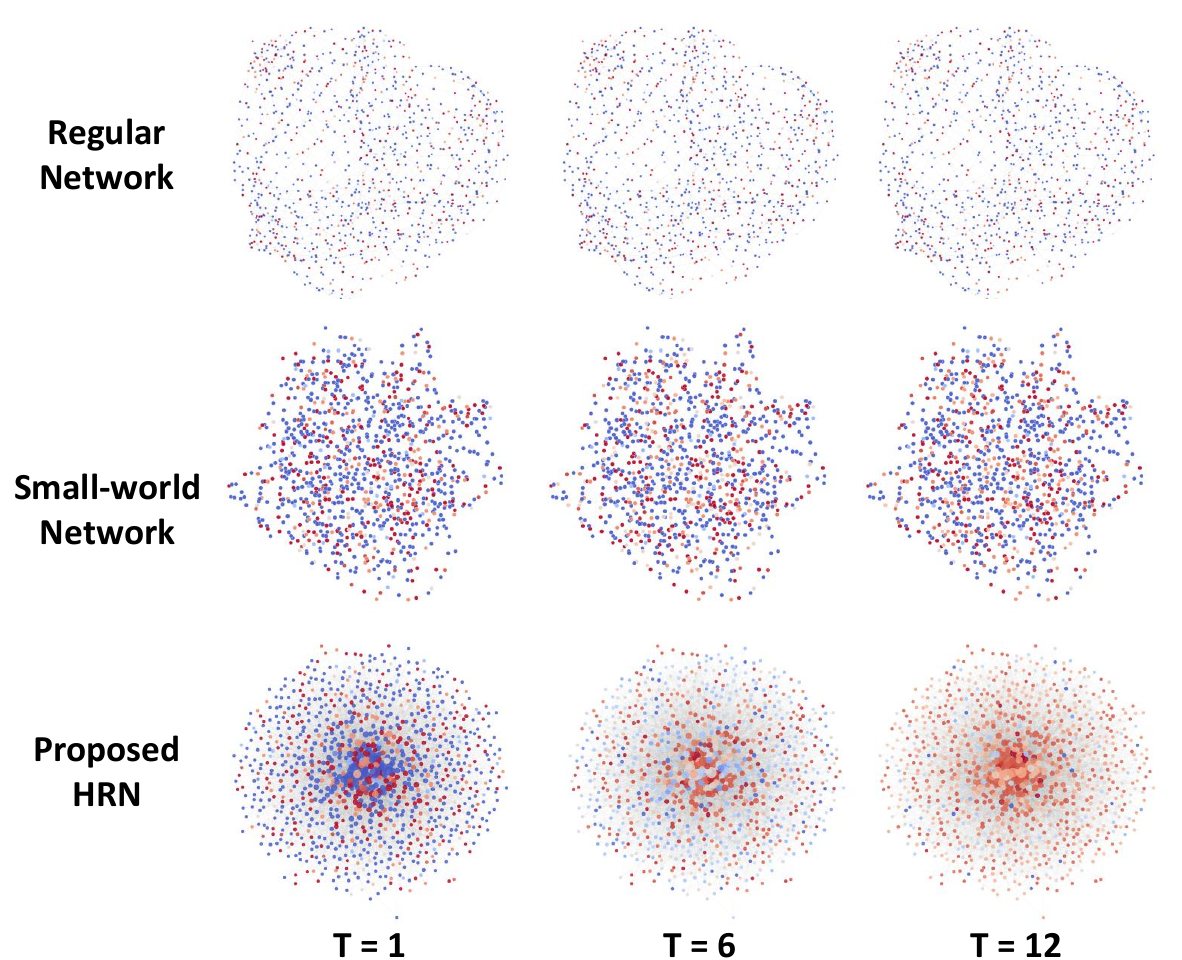}
    \caption{Impact of network structure on opinion dynamics. Colors represent belief intensity from blue at -1 for disbelief to red at +1 for acceptance.}
    \label{fig:topologies}
\end{figure}

\textbf{The Impact of HRN:}
Table~\ref{tab:abl} demonstrates that substituting HRN with alternative topologies consistently degrades fidelity. Notably, even the Small-World baseline underperforms, with $\Delta Bias$ deteriorating to 0.1687. 
Fig.~\ref{fig:topologies} and Table~\ref{tab:topology} (detailed analysis is provided in the Appendix) reveal that these structures suffer from stagnant or delayed propagation due to insufficient clustering coefficients and path lengths. In contrast, HRN accurately captures the explosive spread of rumors, demonstrating its necessity for modeling real-world rumor dynamics.


\textbf{The Impact of CAH:}
Table~\ref{tab:abl} highlights that replacing CAH with static ABM baselines significantly reduces simulation fidelity. The ``-w/o CAH-RA'' variant leads to a 0.0871 increase in $\Delta Bias$ and a 0.2784 rise in $DTW$. Unlike static models, CAH incorporates dynamic susceptibility, allowing agents to adjust influence weights in response to uncertainty. This adaptability is crucial for replicating consensus-seeking behaviors in large-scale rumor propagation.
\begin{figure}[t]
  \centering
  \includegraphics[width=\linewidth]{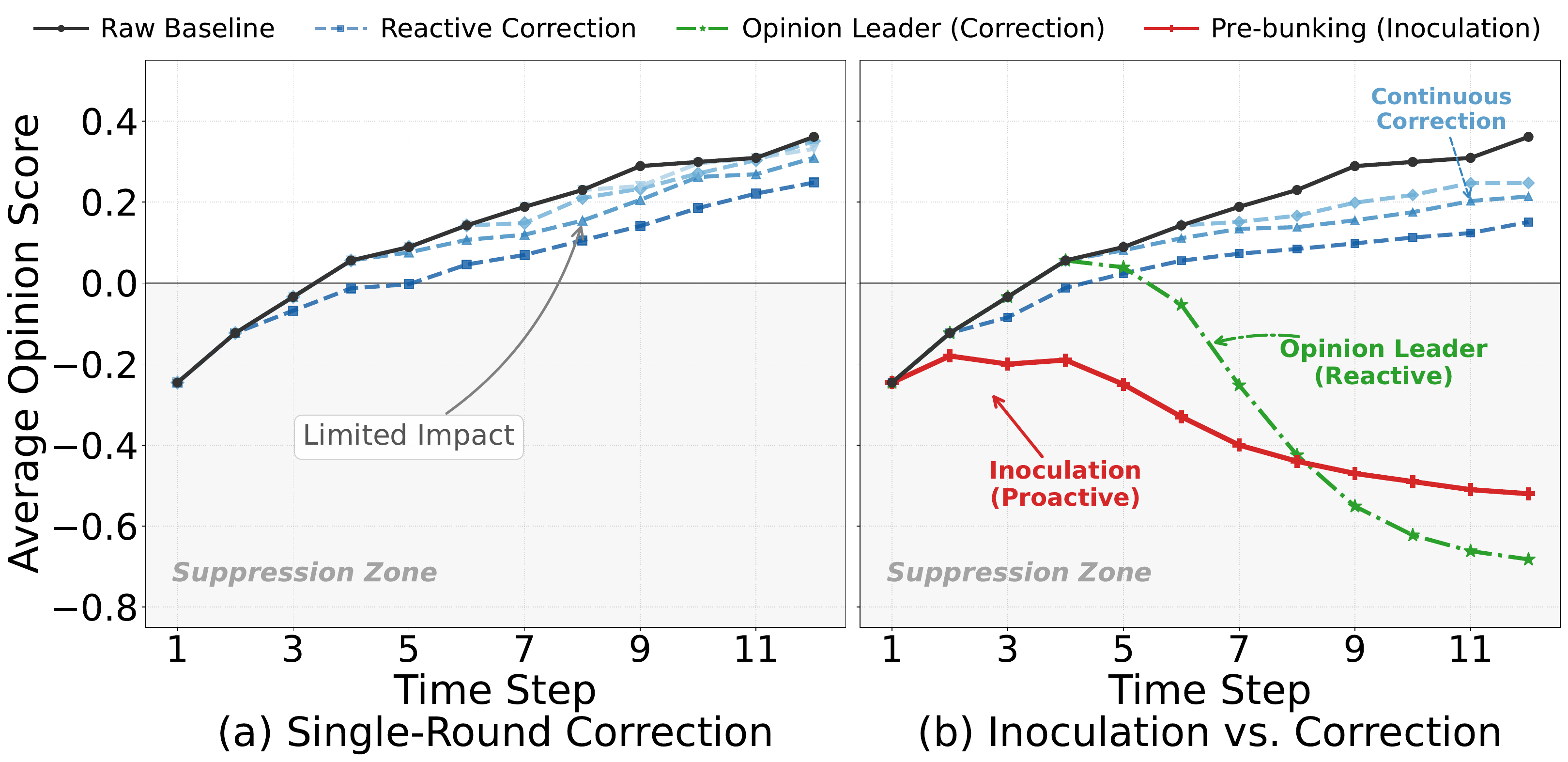}
    \caption{Impact of intervention strategies on rumor dynamics. 
(a) Single-round corrections show limited durability across different start times. 
(b) Proactive Inoculation prevents the outbreak, while Reactive Correction reverses the trend after propagation.}
    \label{fig:intervene}
\end{figure}

\subsection{Intervention Analysis}
We conduct counterfactual analysis on the Xinjiang Cotton dataset to evaluate mitigation paradigms ranging from reactive correction to proactive inoculation (detailed intervention specifications are provided in the Appendix).
As shown in Fig.~\ref{fig:intervene} (a), single-round interventions yield only ephemeral suppression, rapidly succumbing to a ``rebound effect.'' Crucially, Fig.~\ref{fig:intervene} (b) highlights a fundamental mechanism divergence. The Opinion Leader strategy acts as a potent ``corrective antidote,'' achieving sharp consensus reversal even at late stages ($T=6$). In contrast, \textbf{Pre-bunking} induces a ``cognitive inoculation'' effect. By establishing an early \textit{resistance phase}, it flattens the rumor curve with minimal intervention costs, demonstrating that minimizing susceptibility via prevention is strategically superior to post-hoc cure.

\begin{table}[t]
\centering
\resizebox{\columnwidth}{!}{
\setlength{\tabcolsep}{3.5pt}
\begin{tabular}{l|lll}
\toprule
\textbf{Scale} & \textbf{$P_z$} \small{(Polar.)} & \textbf{$GD$} \small{(Disag.)} & \textbf{$NCI$} \small{(Clust.)} \\
\midrule
100       & 0.2248 & 0.1655 & 0.5501 \\
10,000    & 0.3462 \small{($\uparrow$54\%)} & 0.2240 \small{($\uparrow$35\%)} & 0.4372 \small{($\downarrow$21\%)}  \\
1,000,000 & \textbf{0.3854} \small{($\uparrow$\textbf{71\%})} & \textbf{0.2688} \small{($\uparrow$\textbf{62\%})} & \textbf{0.3763} \small{($\downarrow$\textbf{32\%})}  \\
\bottomrule
\end{tabular}
}
\caption{Impact of simulation scale on macro-sociological patterns ($P_z$, $GD$, $NCI$).}
\label{tab:scale}
\end{table}



\subsection{Emergence of Diversity at Scale}
Table~\ref{tab:scale} reveals that complex social phenomena are fundamentally scale-dependent: while small-scale networks of 100 agents collapse into hyper-consensus, expanding to 1,000,000 agents triggers the emergence of polarization. 
Specifically, the 71\% surge in $P_z$ to 0.3854 highlights a significant intensification in global opinion divergence. 
Conversely, the NCI of 0.3763 captures the presence of local cognitive resonance, revealing that agents maintain strong alignment with their neighbors within information cocoons. 
These results demonstrate that the large-scale simulation is indispensable for capturing emergent macro-sociological patterns that remain obscured in smaller simulations.

\begin{figure}[!t]
  \centering
  \includegraphics[width=\linewidth]{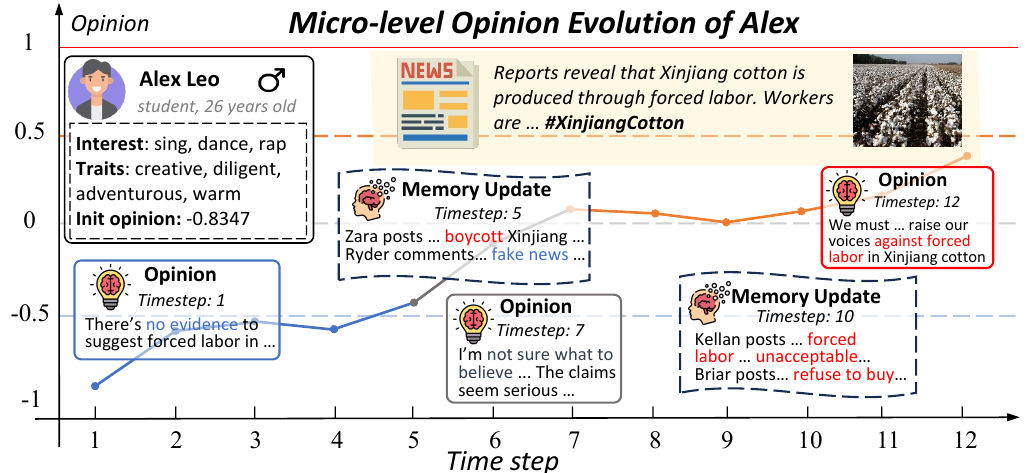}
  \caption{Case study on the spread of the Xinjiang cotton rumor within the small community formed by agent ``Alex'' and its neighboring agents, where their opinions evolved from disbelief to certainty.}
  \label{fig:case}
\end{figure}

\subsection{Case Study}
Fig.~\ref{fig:case} presents a microscopic analysis of opinion evolution by tracking agent Alex, who transitions from initial skepticism to eventual conformity as the rumor permeates his local network. As his memory module records increasing peer endorsements from neighbors such as Zara and Kellan, the resulting cognitive dissonance erodes his disbelief and shifts his stance toward uncertainty by step 7. By step 12, Alex aligns with the dominant local consensus and actively shares supporting content, a progression that mirrors the Spiral of Silence theory \cite{noelle1974spiral}. This trajectory underscores how local social pressure drives individual opinion conformity over time, validating the high fidelity of our agent behavioral modeling.



\section{Conclusion}
This paper introduces RumorSphere, a dynamic and hierarchical resonance framework for million-scale rumor simulation. 
Considering the dynamic role of core agents in rumor evolution, we propose a dynamic interaction strategy that adaptively activates core agents at conflict boundaries via LLMs, enabling million-scale simulations. Furthermore, we design a hierarchical resonance network integrating opinion leaders and local communities to facilitate realistic explosive rumor diffusion.
Extensive experiments demonstrate that RumorSphere significantly outperforms state-of-the-art baselines, achieving an average 26.5\% reduction in simulation bias, providing a powerful tool for understanding and mitigating rumor propagation. 

\section*{Limitations}
Despite the promising performance of RumorSphere, several limitations remain.
First, to reconcile behavioral realism with million-scale scalability, we model the majority of regular agents using the CAH model. Although this strategy effectively captures macro-level dynamics, it inevitably serves as a low-rank approximation of complex human psychology, potentially simplifying fine-grained emotional nuances.
Second, our simulation environment is currently modeled after a single platform architecture (Twitter-like). Real-world rumors often propagate through cross-platform ecosystems with diverse algorithmic mechanisms (e.g., recommendation feeds in TikTok), which fall outside the current scope.
Finally, our experiments focus on textual rumors, leaving the dynamics of multimodal misinformation (e.g., deepfakes, images) as a subject for future investigation.

\section*{Ethical Considerations} 
The development of RumorSphere entails inherent ethical responsibilities, particularly concerning the potential dual-use risks of generative agents. While the framework is designed to elucidate the mechanics of misinformation for defensive purposes, we recognize the risk of misuse in amplifying harmful content. Consequently, we advocate for deploying such simulations strictly within controlled, sandboxed environments equipped with robust content safeguards. Furthermore, data privacy remains a cornerstone of our research; all real-world datasets utilized in this study have undergone rigorous anonymization to ensure confidentiality and prevent the exposure of sensitive personal information. Finally, we emphasize that simulation outcomes should be interpreted transparently as theoretical models, ensuring that this tool contributes constructively to safeguarding the digital information ecosystem.

\bibliography{custom}

\end{document}